\documentclass{aa}  

\usepackage{graphicx}
\usepackage{txfonts}
\usepackage{caption}
\usepackage{subcaption}
\def\av {${A_{V}}$}
\def\deg{$^{\circ}$}
\def\arcsec{$^{\prime\prime}$}
\def\arcmin{$^{\prime}$}
\def\nh {${\rm N_\mathrm{H}}$}

\begin{document}

   \title{4XMM J181330.1-175110: a new supergiant fast X-ray transient}


   \author{M. Marelli\inst{1}
          \and
          L. Sidoli\inst{1}
          \and
          M. Polletta\inst{1,2}
          \and
          A. De Luca\inst{1}
          \and
          R. Salvaterra\inst{1}
          \and
          A. Gargiulo\inst{1}
          }

   \institute{Istituto Nazionale di Astrofisica, Istituto di Astrofisica Spaziale e Fisica Cosmica di Milano, via A. Corti 12, 20133 Milano, Italy\\
             \email{martino.marelli@inaf.it}
             \and
             Department of Physics and Center for Astrophysics and Space Science, University of California at San Diego, 9500 Gilman Drive,
La Jolla, CA 92093, USA
             }

   \abstract
   {Supergiant Fast X-ray Transients (SFXTs) are a sub-class of High Mass X-ray Binaries (HMXBs) in which a compact object accretes part of the clumpy wind of the blue supergiant companion, triggering a series of brief, X-ray flares lasting a few kiloseconds. Currently, only about fifteen SFXTs are known.}
   {The EXTraS catalog provides the timing signatures of every source observed by the EPIC instrument on-board \textit{XMM-Newton}. Among the most peculiar sources, in terms of variability, we identified a new member of the SFXT family: 4XMM J181330.1-17511 (J1813).}
   {We analyzed all publicly available \textit{XMM-Newton}, \textit{Chandra}, \textit{SWIFT}, and \textit{NuSTAR} data pointed at the J1813 position to determine the source’s duty cycle and to provide a comprehensive description of its timing and spectral behavior during its active phase. Additionally, we searched for the optical and infrared counterpart of the X-ray source in public databases and fitted its Spectral Energy Distribution (SED).}
   {The optical-to-MIR SED of J1813 is consistent with a highly-absorbed (A$_V\sim38$) B0 star at $\sim$10 kpc. During its X-ray active phase, the source is characterized by continuous $\sim$thousands seconds-long flares with peak luminosities (2-12 keV) ranging from $10^{34}$ to $4 \times 10^{35}$ erg s$^{-1}$. Its X-ray spectrum is consistent with a high-absorbed power-law model with N$_H \sim 1.8 \times 10^{23}$ cm$^{-2}$ and $\Gamma \sim 1.66$. No spectral variability was observed as a function of time or flux. J1813 is in a quiescent state $\sim$60\% of the time, with an upper-limit luminosity of $8 \times 10^{32}$ erg s$^{-1}$ (at 10 kpc), implying an observed long-term X-ray flux variability $>$500.}
   {The optical counterpart alone indicates J1813 is a HMXB. Its transient nature, duty cycle, the amplitude of observed X-ray variability, the shape and luminosity of the X-ray flares -- and the lack of known X-ray outbursts ($>10^{36}$ erg s$^{-1}$) -- strongly support the identification of J1813 as an SFXT.}


   \maketitle
%

\section{Introduction}

High-mass X-ray binaries (HMXB, \citealt{Kretschmar2019, Fornasini2023} for the most recent reviews) are binary systems composed of a compact object, usually a neutron star (NS), accreting matter from the strong wind of the massive stellar companion. Depending on the spectral type and luminosity class of the donor (either an O- or B-type supergiant or a non-supergiant Be star), HMXBs divide into Supergiant HMXBs (SgXRB, hereafter) and  BeXRBs. 
While in BeXRBs the massive companion rotates fast and expels material via a dense, equatorial decretion disk, in SgXBs the outflowing wind is supersonic and approximately spherical. 
Most BeXRBs \citep{Reig2011} show aperiodic (type II, giant outbursts) and/or periodic (type I outbursts, repeating on the orbital period timescale) luminous X-ray outbursts. During type I outbursts, triggered when the NS is closer to the periastron passage and the Be decretion disk, the peak X-ray luminosity is around 10$^{37}$ erg~s$^{-1}$, while during type II, giant outbursts it is close to Eddington limit (10$^{38}$ erg~s$^{-1}$). While Type I outbursts last about $\sim$20-30\% of the orbital period, Type II ones can cover a number of orbital periods of the system.
In BeXRBs, especially during type II outbursts, the transfer of matter onto the NS forms an accretion disk, while the SgXBs are thought to be wind-fed sources.
SgXRBs \citep{Martinez2017} are equally divided (for number of members) in persistent and transient sources. The latter family is named SFXTs: they were discovered by the INTEGRAL satellite 20 years ago \citep{Sguera2005, Sguera2006, negueruela06} and display extreme properties. They shine at an X-ray luminosity level similar to persistent SgXRBs (a few 10$^{36}$~erg~s$^{-1}$) only during short (lasting hundred to thousand seconds only) flares, with a duty cycle (fraction of time spent in these bright flares) of less than 5\% \citep{Sidoli2018}. These flares are part of shorter (a few days, at most) and less luminous outbursts than BeXRBs. SFXTs are believed to host wind-fed NS, which spend most of their time in a low luminosity state, below 10$^{34}$~erg~s$^{-1}$. Their lowest X-ray luminosity can be as low as a few  10$^{31}$~erg~s$^{-1}$ \citep{Sidoli2021, Sidoli2023}. 

The physical driver (e.g. either the neutron star magnetic field, or its rotational period, or the properties of the captured wind material from the supergiant companion, or the interplay of all them together) of the SFXT phenomenology is still highly debated in the literature. Therefore, the discovery of new members of the class is an important step to define the average behavior of this sub-class and compare the population with predictions of different theoretical models \citep{Bozzo2008, Shakura2013, Shakura2014, Martinez2017, Kretschmar2019}.

A search in the X-ray archives for sources exhibiting pronounced short-term variability is undoubtedly the most promising method for discovering new members of these classes of objects. Among the various X-ray catalogues available, EXTraS stands out as the most powerful tool for such searches. 
The EXTraS project\footnote{Project funded by the EU/FP7 program.} \citep{DeLuca2017} extracted the light curves of all sources observed by the European Photon Imaging Camera (EPIC) aboard the {\it XMM-Newton} spacecraft, which is one of the most sensitive X-ray telescopes ever built, with a large effective area and over 20 years of available data. The main goal of the project is to characterize the variability of these sources. 
In recent years, we have been enhancing the capabilities of EXTraS in terms of both detection and characterization of source variability, extending the analysis to the most recent datasets. A small fraction of unpublished sources exhibiting extreme temporal features (e.g., a high number of Bayesian Blocks or a peculiar shape of the cumulative light curve) have caught our attention.
Among these, we report here on an unidentified source, 4XMM J181330.1-175110 (hereafter J1813), whose transient activity was serendipitously detected by {\it XMM-Newton} during a long observation and subsequently discovered thanks to the EXTraS project.

In this paper we report on the analysis of the available  X-ray (Sect. \ref{sec:obs},\ref{sec:ana}) and optical/mid-infrared data (Sect.~\ref{sec:opt}), leading us to propose an identification with a new high mass X-ray binary (HMXB), in particular with a supergiant fast X-ray transient (SFXT; Sect.~\ref{sec:disc}).

\section{Observations and data reduction} \label{sec:obs}

The sky position of J1813 was serendipitously covered by numerous observations from soft X-ray telescopes due to its proximity to the well-known energetic pulsar PSR J1813-1749 \citep[see e.g.][]{got09}, its pulsar wind nebula, and the associated supernova remnant (approximately 1\farcm 7 away). Given the relative brightness of the nearby pulsar system, this paper utilizes only data sets from X-ray telescopes operating in imaging mode with sufficiently good angular resolution. For instance, we do not consider data from {\it NICER} \citep{gen12}, {\it Suzaku} \citep{mit07}, or Continuous Clocking-mode data from {\it Chandra} \citep{gar03}. Additionally, we do not consider {\it SWIFT/XRT} observations, where J1813 is not detected, and the observations are too short to provide scientific relevance for computing the source's duty cycle, with an upper limit exceeding 5$\times10^{-13}$ erg cm$^{-2}$ s$^{-1}$. It is worth noting that these short observations cover a total exposure time of 8 ks and therefore are negligible. Table 1 presents the observations used in this paper along with their main characteristics.
As a final consideration, the {\it XMM-Newton} and {\it SWIFT/XRT} observations are contaminated by relatively faint straylight annuli from the Low-Mass X-ray Binary GX 13+1, located 45\arcmin\ away; however, this contamination does not affect the position of J1813 or any of the analyses presented in this paper.

The most accurate X-ray position of J1813 is derived from the {\it Chandra} Source Catalog (CSC) 2.1 \citep{eva24}, located at R.A.(J2000.0)= 18$^h$13$^m$30$^s$.17, Dec=$-$17$^{\circ}$51$^{\prime}$10$^{\prime\prime}$.3 (with a 0.30$^{\prime\prime}$ 1$\sigma$ statistical plus systematic error). This position is consistent with that from the {\it XMM-Newton} Source Catalog 3XMM-DR14 \citep{web20}.

\begin{table}
\centering
\caption{X-ray observations log
}
\resizebox{\columnwidth}{!}{
\begin{tabular}{cccc}
\hline
\hline
Satellite & ObsID & Exposure\tablefootmark{a} & Detected$^a$\\
 &  & (ks) & 10$^{-13}$ erg cm$^{-2}$ s$^{-1}$\\
\hline
{\it XMM-Newton} & 0302450201 & 16 & $<0.77^b$\\
{\it XMM-Newton} & 0552790101 & 99 & Y\\
{\it Chandra} & 6685 & 30 & Y\\
{\it Chandra} & 17440 & 17 & Y\\
{\it Chandra} & 17695 & 13 & Y\\
{\it SWIFT$^{c}$} & 00035115002 & 7 & $<$3\\
{\it SWIFT$^{c}$} & 03110848001 & 5 & $<$4\\
{\it SWIFT$^{c}$} & 00035115001 & 3 & $<$5\\
{\it NuSTAR} & 30364003002 & 29 & $<0.73^d$\\
\hline
\hline
\end{tabular}
}
\tablefoot{
Log of the X-ray observations.\\
\tablefoottext{a}{Detection or the 3$\sigma$ upper limit unabsorbed flux in the 2-10 keV band, using the spectrum described in Section \ref{sec:spec}.}\\
\tablefoottext{b}{For this upper limit, we conservatively used only the pn camera in the 2-12 keV energy range; no proton flares were detected during this observation.}\\
\tablefoottext{c}{{\it SWIFT/XRT} upper limits and detections were performed using respectively \footnote{\url{https://www.swift.ac.uk/2SXPS/ulserv.php}} and \footnote{\url{https://www.swift.ac.uk/user_objects/}} \citep{eva20}, when available, using {\tt HEASOFT} v.6.32.}\\
\tablefoottext{d}{For this upper limit, we combined the FPMA and FPMB cameras using good quality events fits file in the 2-10 keV energy band. We note that the source area is contaminated by the nearby PSR J1813-1749 system.}
}
\label{tab:XrayObs}
\end{table}

For most of the X-ray analysis in this work, we used {\tt HEASoft} \citep{hea14} v.6.34, CIAO \citep{fru06} v.4.16, and SAS \citep{gab04} v.20.0.0.\\
For the {\it XMM-Newton} analysis, the event files were barycentered using the SAS tool {\tt barycen}. Proton flares did not significantly affect the analyzed observations. We extracted source events from a circle with a radius of 25$^{\prime\prime}$ centered on the source position and background events from a nearby 50$^{\prime\prime}$-radius circle, also with similar contamination from the nearby PSR J1813-1749 nebula. It is noteworthy that in the longest observation, 0552790101, which was the only one with a detection, the position of J1813 was covered only by MOS cameras.\\
For the MOS spectral analysis of this observation, we extracted photons in the 0.3-12 keV energy range, with  patterns 0-12, and only during the 'active' phase (see Section \ref{sec:lc}).
The {\it XMM-Newton} temporal analysis was carried out using the EXTraS tools \citep[][Marelli et al. in preparation]{del21} in the 2-12 keV energy band due to its highly absorbed spectrum (see Section \ref{sec:spec}). EXTraS light curves from the two MOS cameras were binned using the same time bins (beginning with the zero XMM-Newton reference time), and therefore the count rate of each bin was converted to flux using the best-fitted spectrum (see Section \ref{sec:spec}). Finally, we calculated the weighted mean, per bin, of the flux curves from MOS1/2 to obtain the final light curve, as previously done in \citet{mar17,mar18,del20,del22}.\\
For the {\it Chandra} analysis, we extracted source events from a circle with a radius of 2$^{\prime\prime}$ centered on the source position and background events from a nearby 20$^{\prime\prime}$-radius circle with similar contamination from the nearby PSR J1813-1749 system. Again, we extracted photons in the 2-10 keV energy range for the {\it Chandra} timing analysis of J1813.

\section{X-ray Data Analysis} \label{sec:ana}

\subsection{Temporal Analysis} \label{sec:lc}

\begin{figure}
\centering
\includegraphics[width=0.49\textwidth]{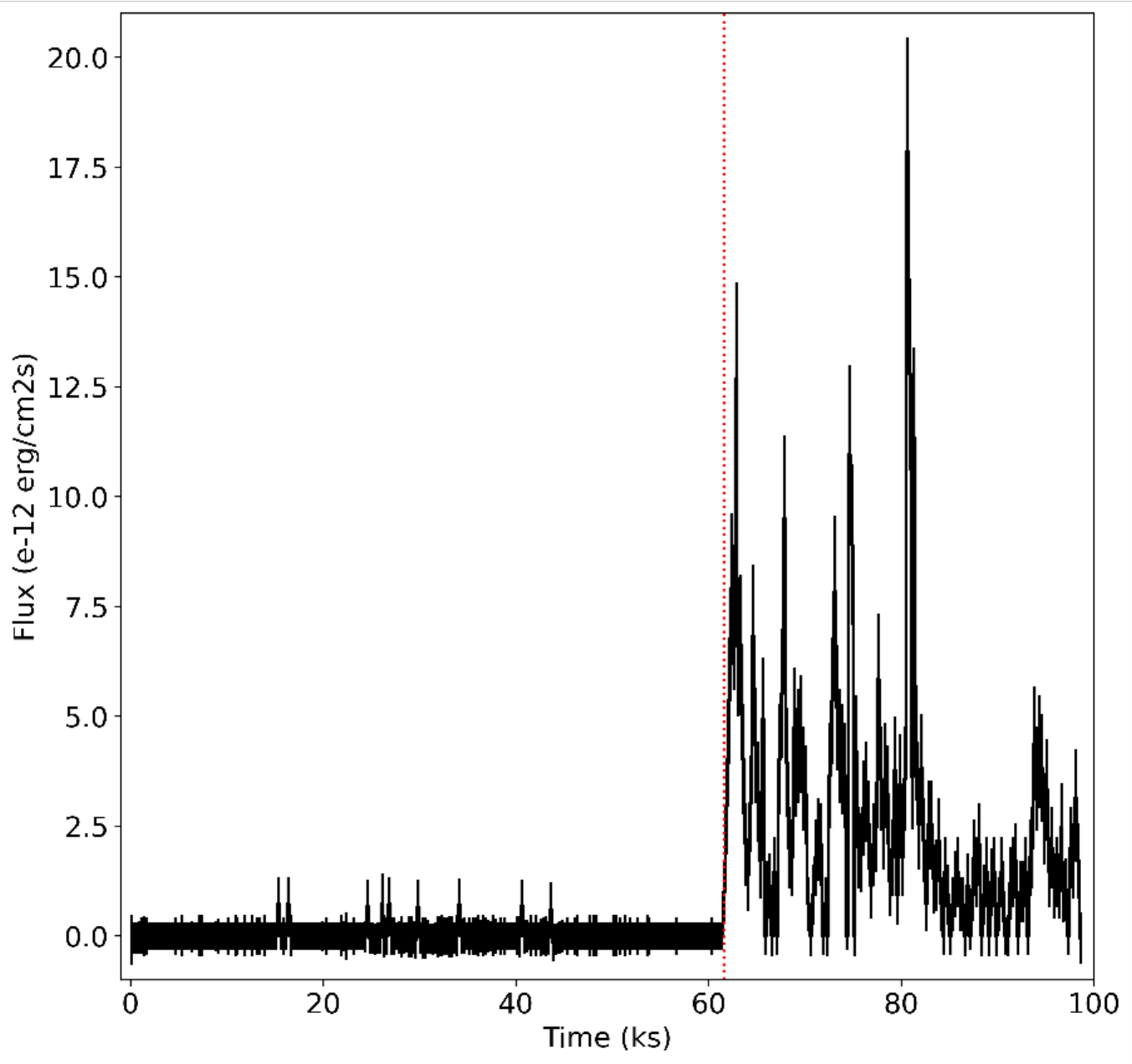}
\caption{EXTraS light curve of J1813 in the 2-12 keV energy band, 250s time bin, T0=354548500. The dotted, red line marks the end of the quiescent phase.}
\label{fig:lctot}
\end{figure}

As described in Section \ref{sec:obs}, we constructed the 2-12 keV {\it XMM-Newton} flux curve of J1813 (obs. id 0552790101) using EXTraS tools and taking into account the contributions from the two MOS cameras (Figure \ref{fig:lctot}). During the first 61.5 ks of observation (before T0 = MJD 51224.39856 = 35458500 s, in {\it XMM-Newton} time), the source was in a quiescent state and compatible with no flux. For the remaining 37.5 ks of the observation, the curve exhibits an apparent and bright flaring behavior, with the brightest flare reaching $\sim 2 \times 10^{-11}$ erg cm$^{-2}$ s$^{-1}$ (assuming the spectral fit derived in Section \ref{sec:spec}).\\

\begin{figure*}
\centering
\includegraphics[width=\textwidth]{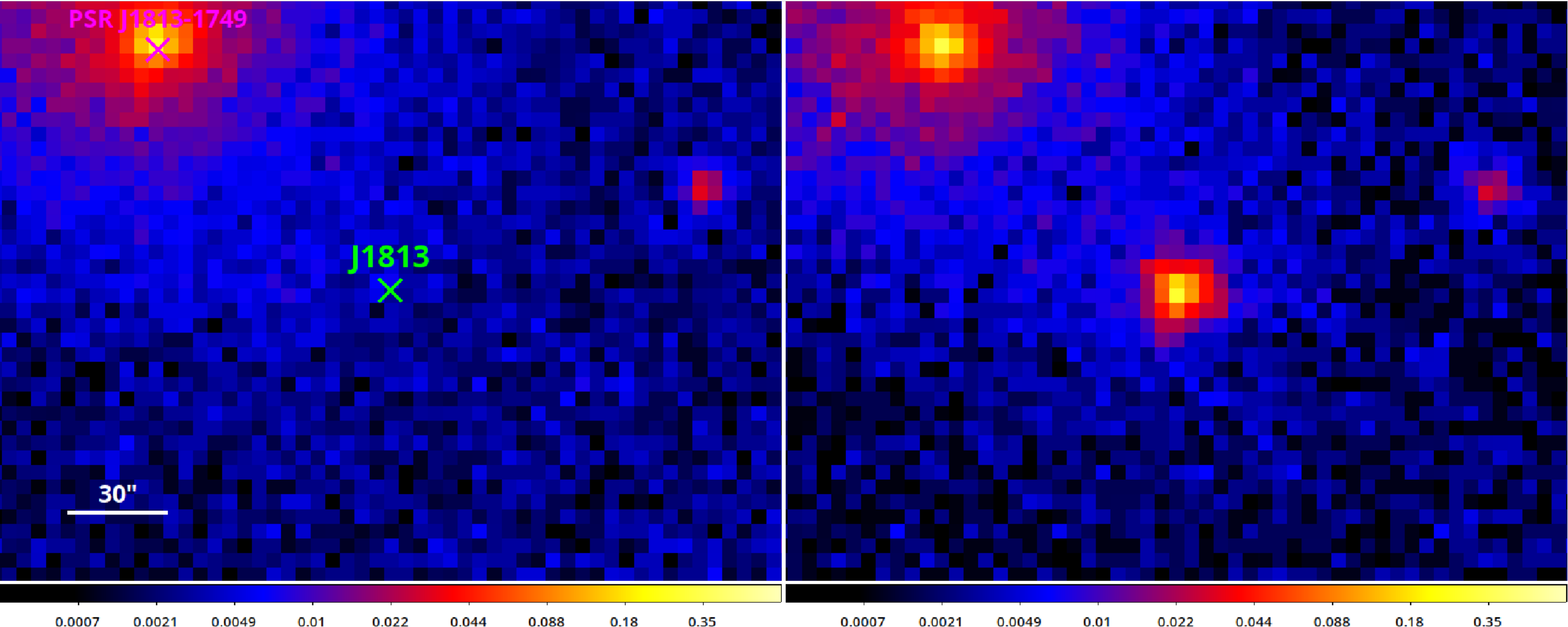}
\caption{{\it XMM-Newton} exposure-corrected image of J1813 (2-12 keV energy band, observation 0552790101) in counts ks$^{-1}$ arcsec$^{-2}$. {\it Left Panel:} quiescent state, with T<354610080; {\it Right Panel:} active state, with T>354610080.}
\label{fig:image}
\end{figure*}

Focusing on the quiescent state of J1813, we extracted the 2-10 keV {\it XMM-Newton} image, pattern 0-12, from the two MOS cameras, as shown in Figure \ref{fig:image}. The source detection procedure using the SAS tool {\tt edetect chain}, performed in a single band with both {\it XMM-Newton} cameras, does not detect J1813. We computed the 3$\sigma$ upper limit using the definition of signal-to-noise and a source-free region, conservatively for a single camera, yielding an upper limit count rate of $\sim 10^{-3}$ counts s$^{-1}$. This is in agreement with the count rate of the faintest source detected at $>3\sigma$ by {\tt edetect chain}, which is $8 \pm 2 \times 10^{-3}$ counts s$^{-1}$. Using the source spectrum from Section \ref{sec:spec}, we obtain a 3$\sigma$ upper limit on the absorbed flux (2-12 keV) for the quiescent state of $\sim 5 \times 10^{-14}$ erg cm$^{-2}$ s$^{-1}$.

\begin{figure*}
\centering
\includegraphics[width=0.49\textwidth]{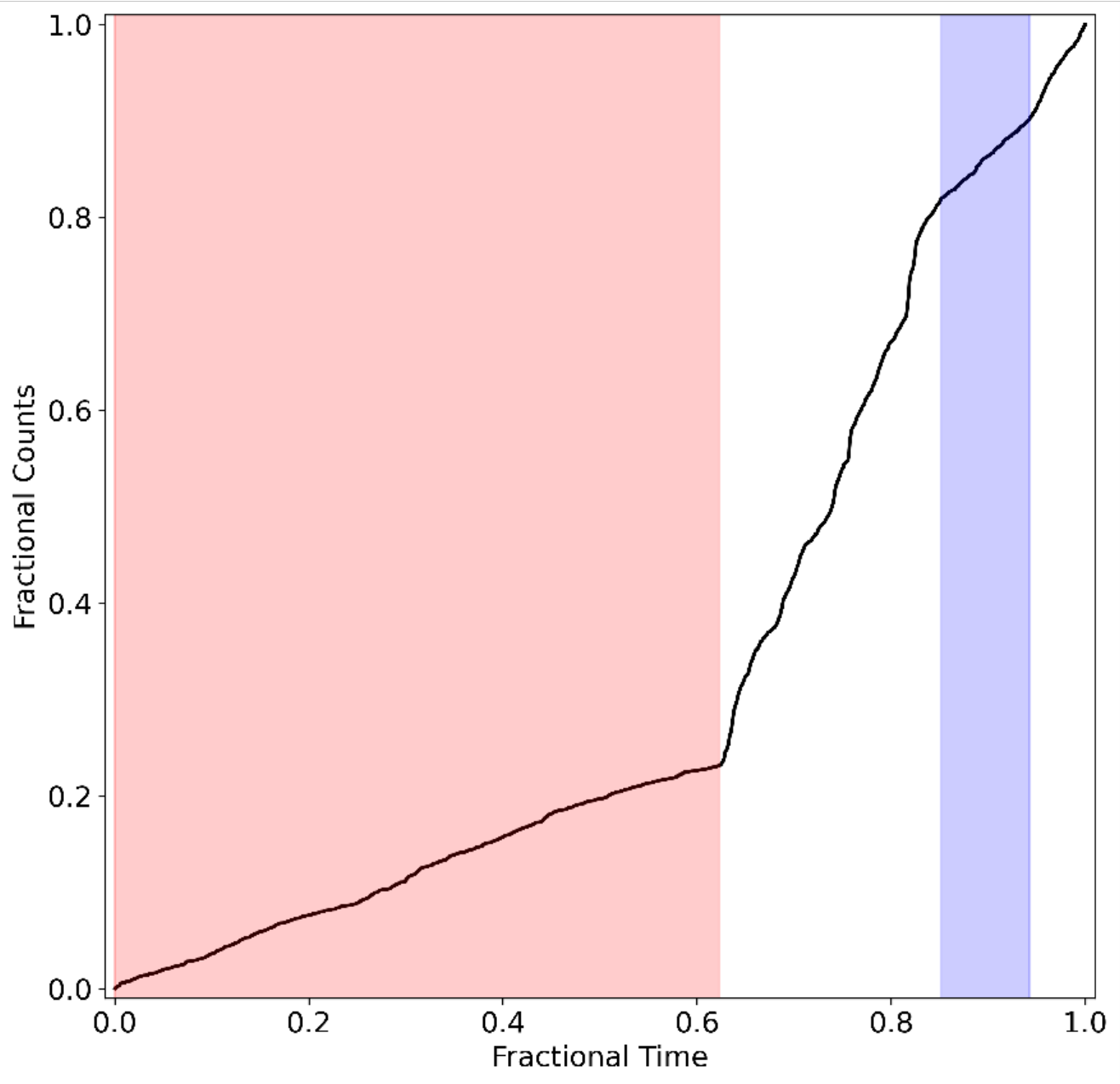}
\includegraphics[width=0.49\textwidth]{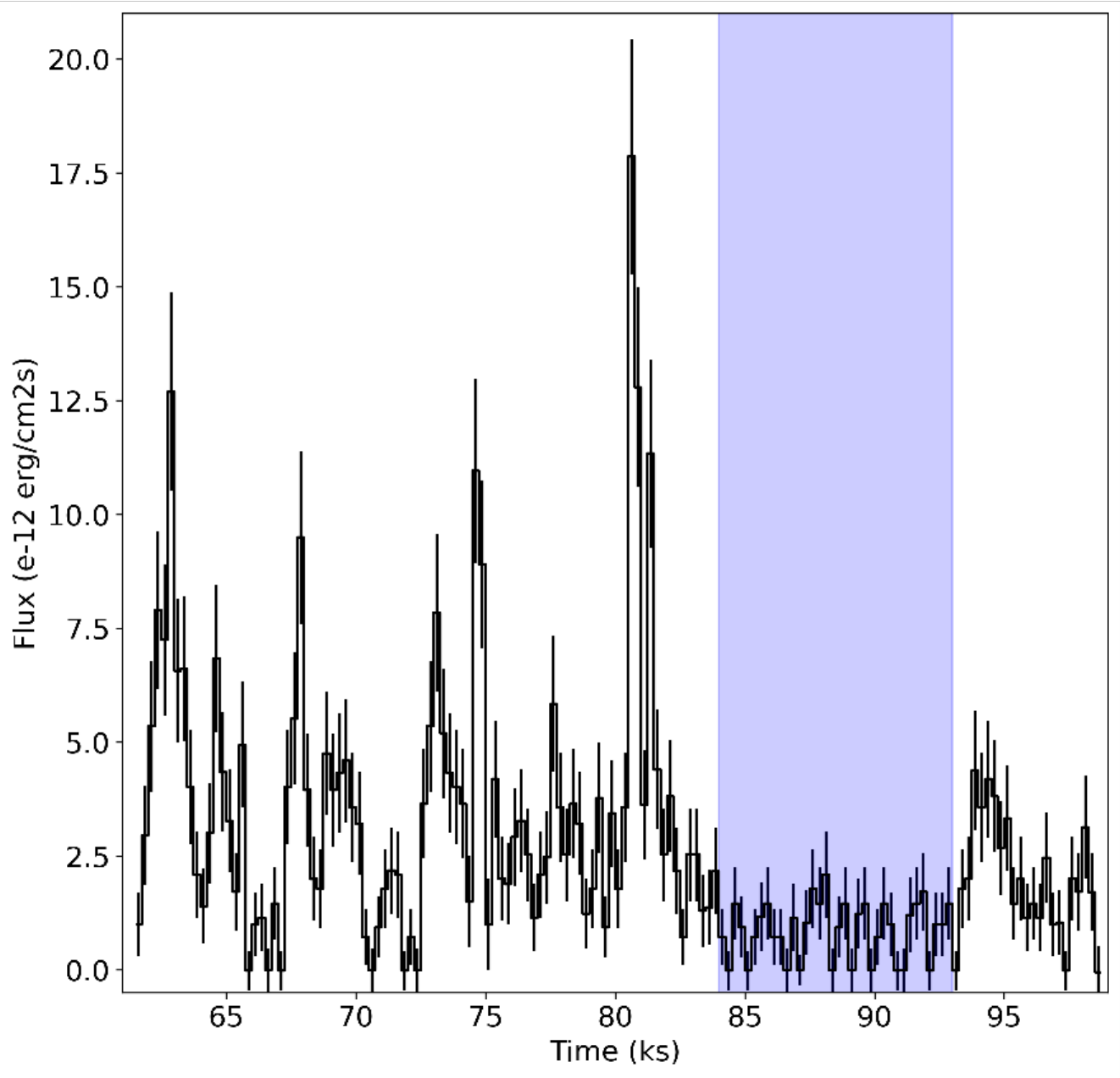}
\caption{{\it Left Panel:} cumulative curve of the events taken from the {\it XMM-Newton} region around J1813 in the 2-12 keV energy band and the cuts described in Section \ref{sec:obs} (we note that it's not background-subtracted). The red area marks the quiescent state; the blue area marks the possible active, not-flaring "intermediate" state. {\it Right Panel:} we show the zoom of the EXTraS flux curve, shown in Figure \ref{fig:lctot}, with only the active part of the curve (T0+61500 s). The blue area is the same as in the left panel.}
\label{fig:lc_final}
\end{figure*}

\begin{figure}
\centering
\begin{subfigure}[t]{0.5\textwidth}
\centering
\includegraphics[width=\textwidth]{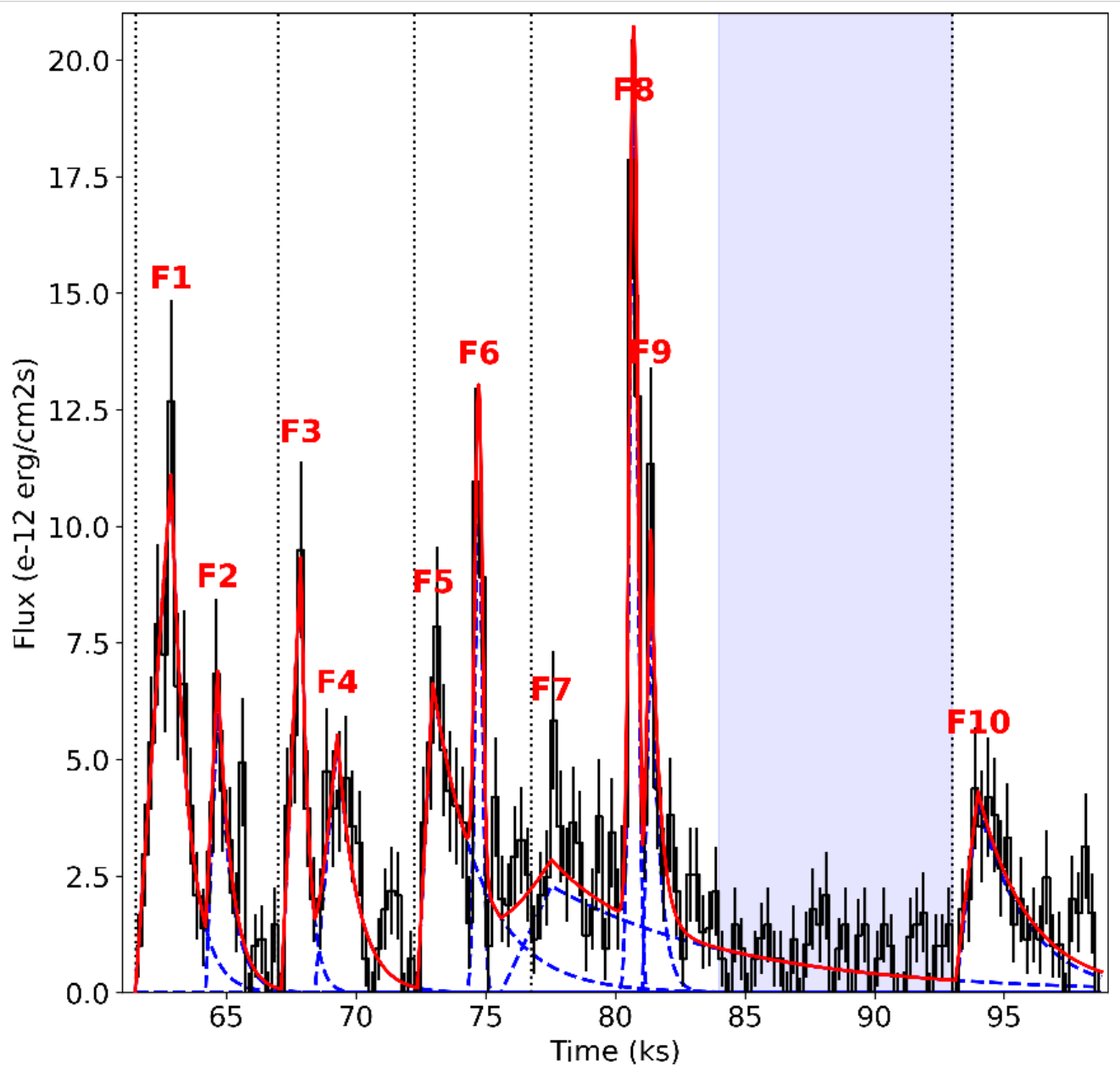} 
\end{subfigure}

\begin{subfigure}[t]{0.5\textwidth}
\centering
\includegraphics[width=\textwidth]{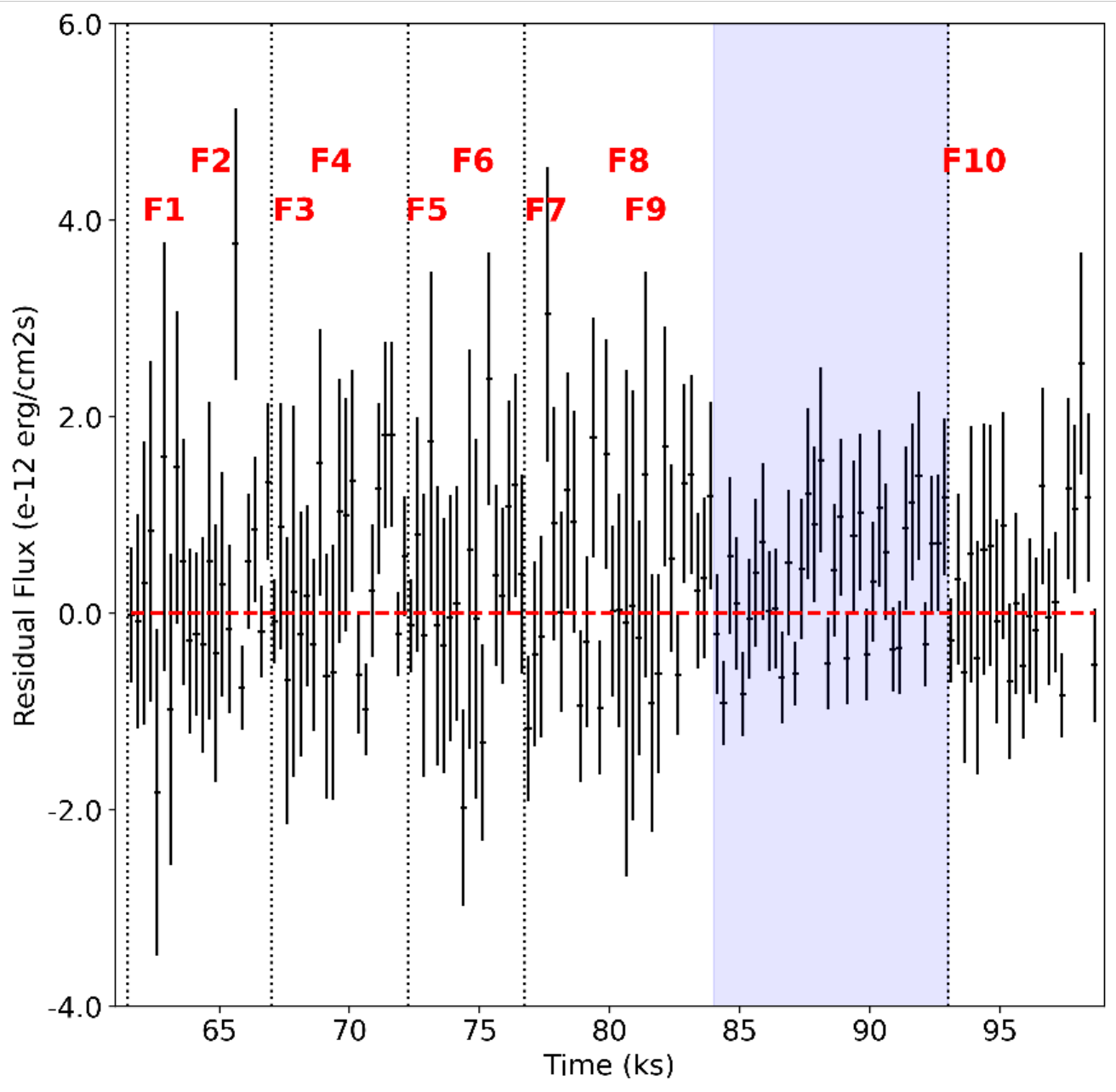} 
\end{subfigure}

 \caption{We show the active part of the {\it XMM-Newton} EXTraS flux curve (from T0+61500 s), as in Figure \ref{fig:lc_final}, right panel. {\it Upper Panel:} we report the flux curve (in black) plus the best-fitted model described in Section \ref{sec:lc} (in red), that is composed by a series of ten different flares (in blue), {\it Lower Panel:} we show the residual curve of the model reported in the Upper panel.}
\label{fig:model}
\end{figure}

\begin{table}
\centering
\caption{Parameters of the flux curve fitting 
}
\resizebox{\columnwidth}{!}{
\begin{tabular}{cccccc}
\hline
\hline
 & L$_{max}^{a}$ & T$_{max}$ & T$_{dec}$ & T$_{ris}$ & L$_{tot}^{a}$\\
 & (10$^{34}$ erg s$^{-1}$) & (ks) & (s) & (s) & (10$^{38}$ erg)\\
\hline
1 & 22.8$\pm$3.1 & 62.9$\pm$0.1 & 646$\pm$205 & 1377$\pm$195 & 30.5\\
2 & 12.7$\pm$3.4 & 64.6$\pm$0.1 & 527$\pm$187 & 468$\pm$273 & 9.6\\
3 & 19.2$\pm$5.3 & 67.9$\pm$0.2 & 303$\pm$249 & 755$\pm$201 & 13.0\\
4 & 11.1$\pm$2.4 & 69.3$\pm$0.2 & 778$\pm$210 & 869$\pm$413 & 13.5\\
5 & 13.5$\pm$2.7 & 73.0$\pm$0.2 & 1857$\pm$1019 & 598$\pm$176 & 29.1\\
6$^b$ & 21.6$\pm$5.3 & 74.74$\pm$0.04 & - & 200$\pm$114 & 7.5\\
7 & 4.8$\pm$1.4 & 77.6$\pm$0.6 & 7216$\pm$2010 & 1946$\pm$1279 & 38.4\\
8$^b$ & 39.3$\pm$0.5 & 80.71$\pm$0.02 & - & 221$\pm$37 & 15.4\\
9 & 17.5$\pm$6.3 & 81.4$\pm$0.3 & 326$\pm$199 & 317$\pm$799 & 8.4\\
10 & 8.4$\pm$1.7 & 94.0$\pm$0.3 & 1883$\pm$526 & 863$\pm$561 & 19.4\\

\hline
\hline
\end{tabular}
}
\tablefoot{
Parameters of the best-fitted model for the active part of the {\it XMM-Newton} EXTraS flux curve. Each row reports the parameters of a different FRED component, the maximum luminosity, the time of the maximum (after T0+61.5 ks), the exponential decay time, the linear rising time and the integral of the model component, respectively. This fit is statistically acceptable, with $\chi^2$=146.9, 111 d.o.f. and a nhp=1.3$\times10^{-2}$.\\
\tablefoottext{a}{Assuming a 10 kpc distance.}\\
\tablefoottext{b}{Here, we fitted a Gaussian. We report the 2-12 keV maximum luminosity, the time of the maximum and the sigma of the Gaussian, respectively.}\\
}
\label{tab:lcpar}
\end{table}

During the active state, the temporal behavior of J1813 resembles a series of FRED (Fast Rise, Exponential Decay) flares, with maximum fluxes ranging from $\sim 2 \times 10^{-12}$ erg cm$^{-2}$ s$^{-1}$ to $\sim 2 \times 10^{-11}$ erg cm$^{-2}$ s$^{-1}$.
Figure \ref{fig:lc_final} (left panel) shows the cumulative light curve of events from the source region. This representation clearly marks the onset of the active state at t = T$_0$ + 61.5 ks. Additionally, there is some indication of a different not-flaring active "intermediate" state between T$_0$ + 84 ks and T$_0$ + 93 ks, where the slope appears uniform and lies between the quiescent and flaring states. Figure \ref{fig:lc_final} (right panel) presents the flux curve during the active state, which remains roughly constant and non-zero. A constant fit of the "intermediate state" (in blue in Figure \ref{fig:lc_final}) provides an acceptable model (null hypothesis probability, nhp = 0.30) with a flux of $5.3 \pm 1.0 \times 10^{-13}$ erg cm$^{-2}$ s$^{-1}$.
However, we cannot exclude the possibility that this state consists of one or more small flares (similar to those observed in the {\it Chandra} data) and/or represents the tail end of previous flares.\\
To investigate this further, we attempted to fit the flux curve of the active state with a series of FRED flares. As shown in Figure \ref{fig:lc_final}, most of the flares are not isolated. Therefore, we created a segmentation of data with Scargle’s Bayesian Blocks \citep{sca13}, using the same events we used to create the cumulative. In this case, the lack of the background subtraction is not relevant since it is almost constant. This resulted in curve with five (single or double-peaked) separated bumps. Thus, we divided the EXTraS curve into the five corresponding time intervals (denoted by dotted vertical lines in Figure \ref{fig:model}), each one containing a bump in the Bayesian Blocks curve. For each interval, we fitted one or more FRED models, using the fewest possible number of components to achieve an acceptable ($3\sigma$) fit. The only exceptions were the 6th and 8th flares: due to their sharpness, we used a simpler Gaussian model instead. This analysis resulted in the identification of ten distinct flares.
Finally, using the best-fitting parameters from each interval as starting points, we performed a simultaneous fit with 38 free parameters. The resulting fit is statistically acceptable, with $\chi^2$ = 146.9 for 111 degrees of freedom, which corresponds to a nhp of $1.3 \times 10^{-2}$. This appears to be the minimum number of flares required to adequately ($3\sigma$) fit the active part of the {\it XMM-Newton} flux curve. Table \ref{tab:lcpar} lists the best-fitting parameters, and Figure \ref{fig:model} displays the corresponding fit.
Since most of the flares are well defined, with at most a minor contamination from the others, there is no degeneration in the parameters and each flare is well constrained. The only exception is flare 7: basically, this flare is necessary to represent the not-flaring "intermediate" state between T$_0$ + 84 ks and T$_0$ + 93 ks, it's not well-constrained (e.g. it can be replaced by a constant term during that period, with variations of the nearby FRED flares) and it has parameters quite different to the others. So, although the entire active state can be described by a series of FRED, we cannot exclude that flare 7 is the sum of two or more shorter flares and/or the "constant" term that dominates in that peculiar period.

\begin{figure*}
\centering
\includegraphics[width=0.33\textwidth]{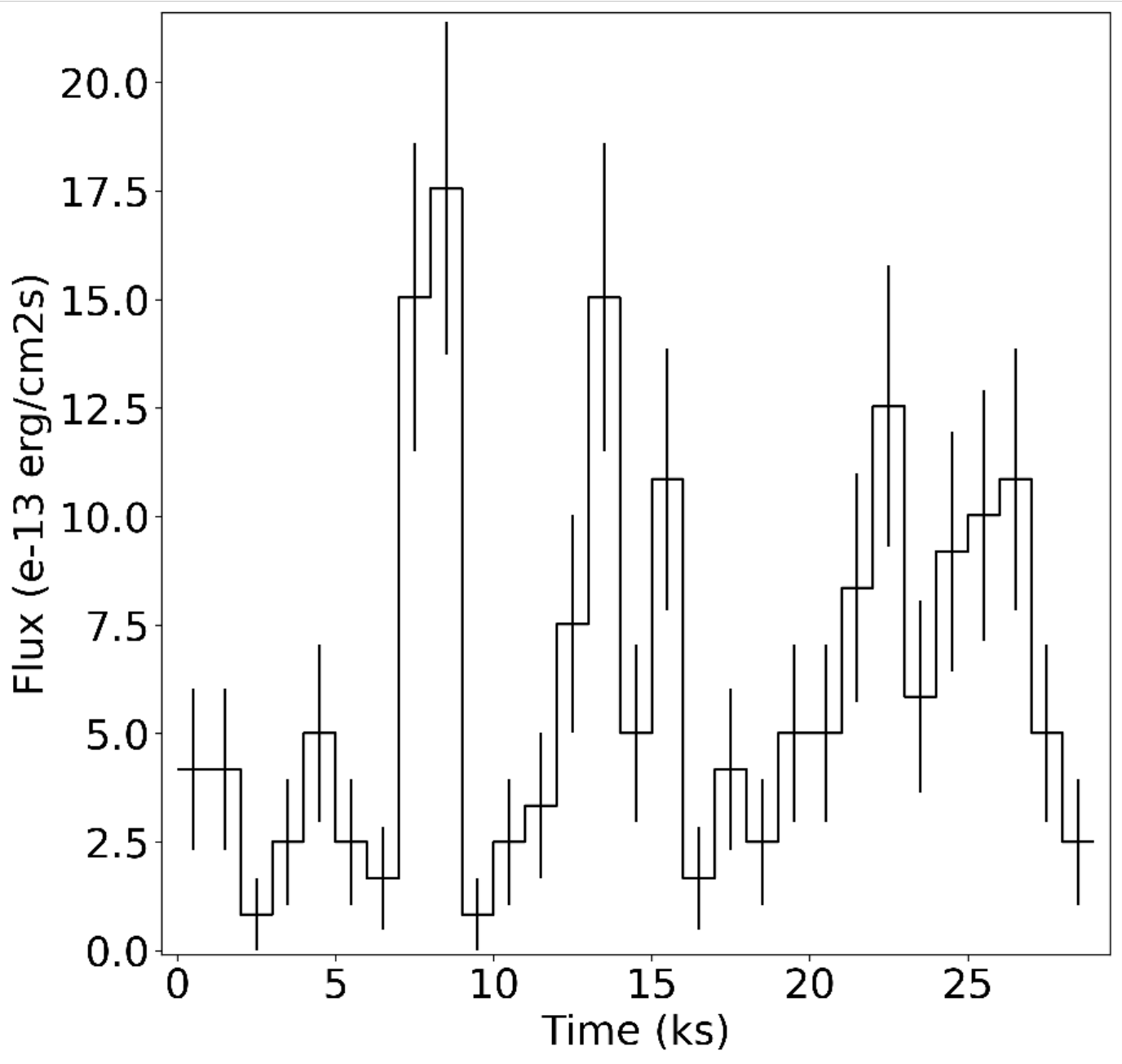}
\includegraphics[width=0.33\textwidth]{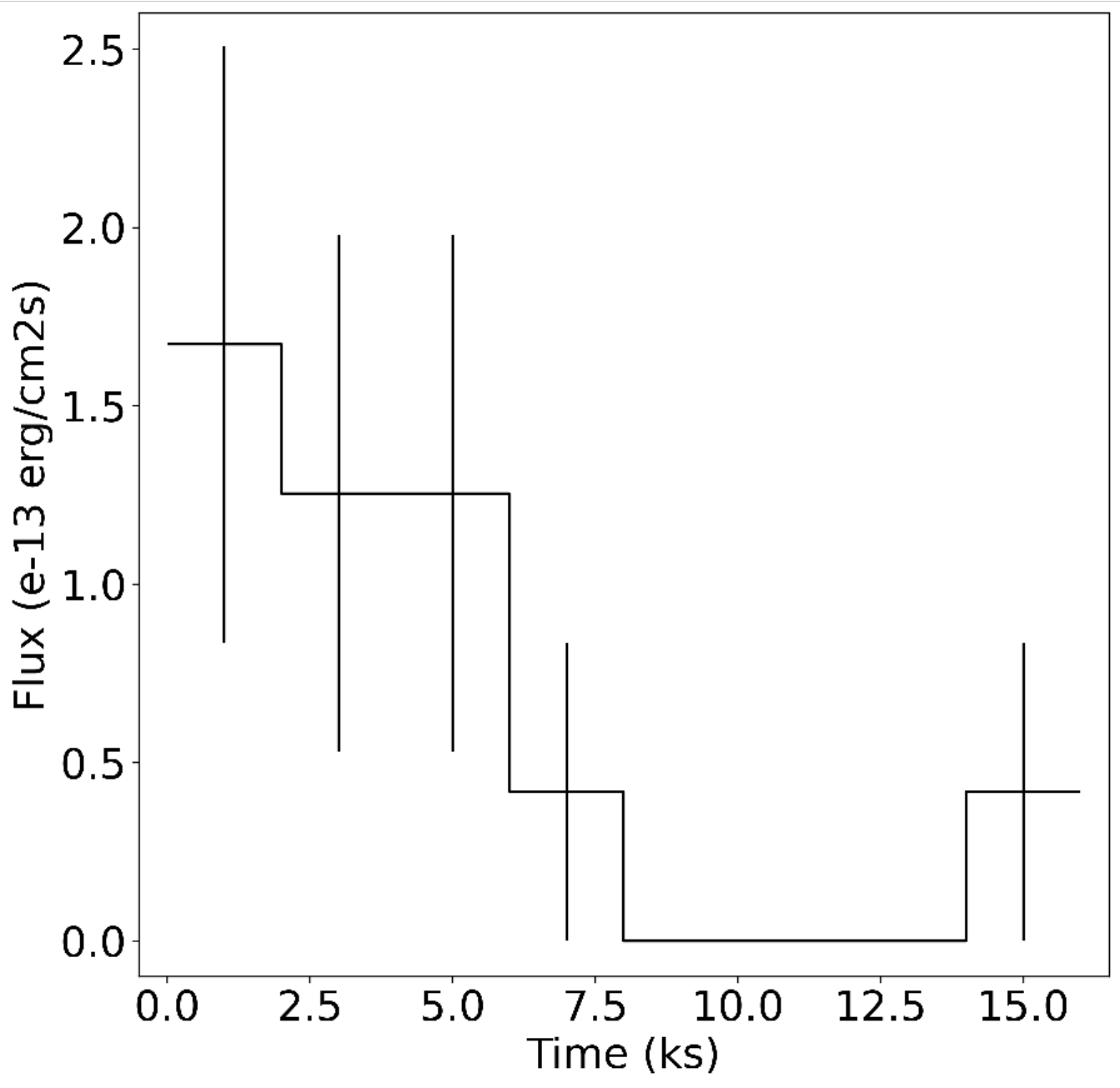}
\includegraphics[width=0.33\textwidth]{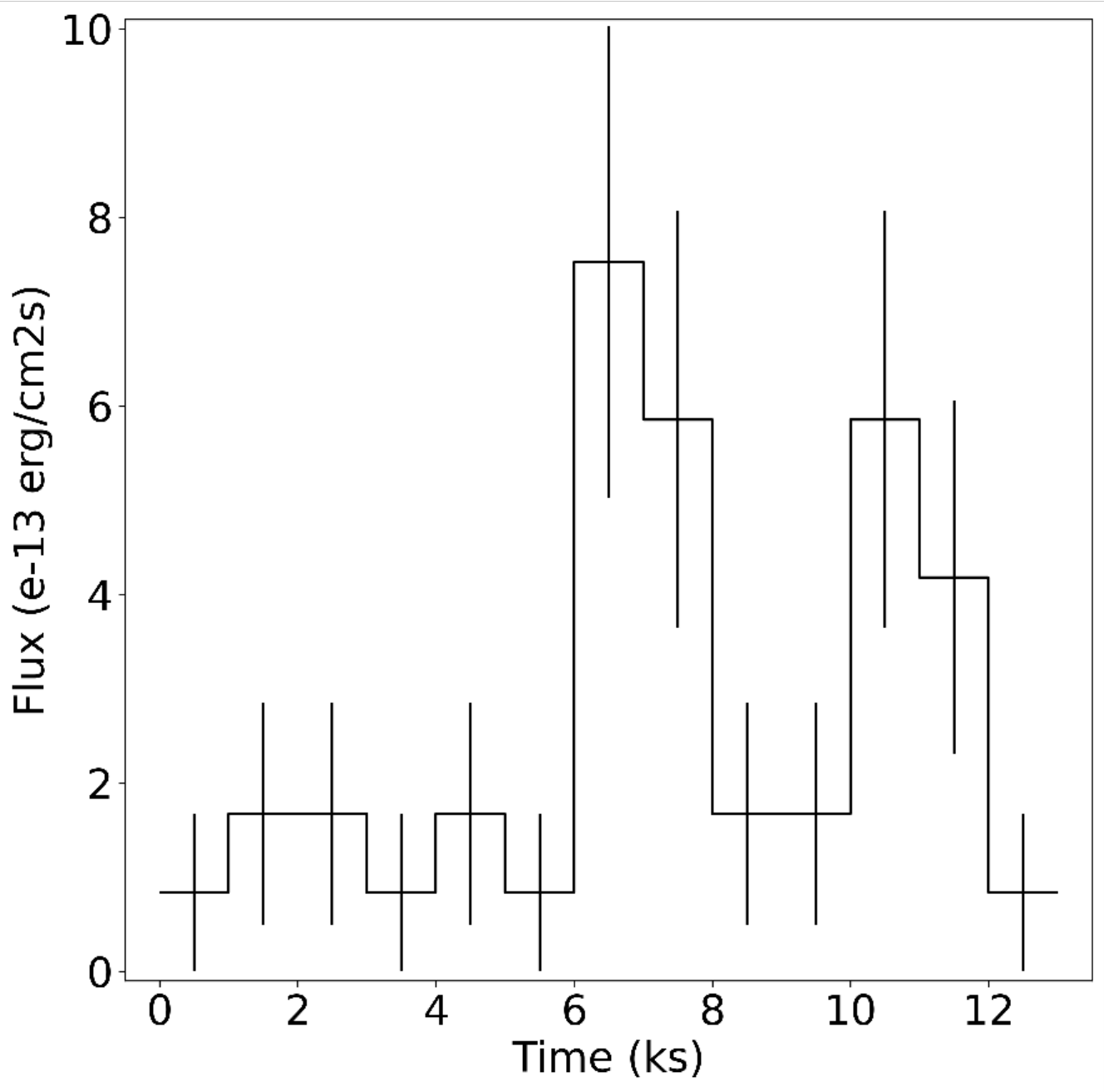}
\caption{{\it Chandra} flux curves taken in 2-10 keV energy range. The count rate was converted into 2-12 keV flux using the best-fitted spectrum (see Section\ref{sec:spec}). {\it Left Panel:} observation 6685, 1ks time bin, T0=274669518; {\it Central Panel:} observation 17440, 2ks time bin, T0=581548320; {\it Right Panel:} observation 17695, 1ks time bin, T0=580947681.}
\label{fig:lc_chandra}
\end{figure*}

The {\it Chandra} flux curves (Figure \ref{fig:lc_chandra}) display a behavior similar to the {\it XMM-Newton} one: two of them consist of a series of FRED-like flares with maximum fluxes ranging from $\sim 5 \times 10^{-13}$ erg cm$^{-2}$ s$^{-1}$ to $\sim 2 \times 10^{-12}$ erg cm$^{-2}$ s$^{-1}$. In the remaining one, observation id 17440, we observe a decrease in flux (compatible with the end of a flare) followed by $\sim$10 ks of quiescence -- compatible (within $1\sigma$) with the noise, and incompatible ($3\sigma$) with the first part of the observation. A source detection during this last period using the CIAO tool {\tt wavdetect} does not yield a detection at the position of J1813. Applying the same method as for {\it XMM-Newton}, we obtain a 3$\sigma$ upper limit of $\sim 8 \times 10^{-4}$ counts s$^{-1}$. This translates into the lowest upper limit on the absorbed flux (2-12 keV) for the quiescent state of $\sim 4 \times 10^{-14}$ erg cm$^{-2}$ s$^{-1}$ (a luminosity of 8 $\times 10^{32}$ erg s$^{-1}$ at 10 kpc).\\

Taking into account the observations reported in Section \ref{sec:obs} and the considerations above, we can compute the duty cycle of J1813, i.e. the percentage of time spent by the source above a certain flux. Over 219 ks of total exposure, the source was in an active state (i.e., a flaring state with a flux of at least a few $10^{-13}$ erg cm$^{-2}$ s$^{-1}$ up to a few $10^{-11}$ erg cm$^{-2}$ s$^{-1}$) for approximately 87.5 ks, which is about 40\% of the time. During the remaining 60\% of the time, the source was in a quiescent state, with a 3$\sigma$ upper limit flux of $4 \times 10^{-14}$ erg cm$^{-2}$ s$^{-1}$.

Finally, we investigated the hard X-ray band searching for possible bursts. No transient hard X-ray source (20-100 keV) has been reported at the J1813 position. We note that J1813 is located only 1.7 arcmin offset from the hard X-ray source IGR J18135-1751, a persistent source (SNR, PSR, PWN) showing a flux of 1.4 mCrab (about 1.1$\times$10$^{-11}$ erg cm$^{-2}$ s$^{-1}$; 20-40 keV; \citealt{Bird2016}). Its positional error radius  with IBIS/ISGRI is 1.4 arcmin (90\% c.l.) and its proximity to J1813 hampers any detection of J1813 with {\it INTEGRAL}, at similar hard X-ray fluxes. In case of an SFXT outburst, with a much brighter (ten times or larger) flux (i.e. with a luminosity of about 10$^{36}$ erg s$^{-1}$, as usually shown by SFXTs at their flare peak), it might have been detected (although proper simulations would be needed to quantify this statement). However, no such activity has ever been reported in the literature, to the best of our knowledge. It is worth mentioning that \citet{Bird2016} catalog performs also a search for transient sources on many different timescales in {\it INTEGRAL} data, reporting no hard X-ray sources at the {\it XMM-Newton} position. After 2016, no publication or telegram have ever reported any transient source consistent with J1813. In the {\it SWIFT/BAT} catalogs the only detected source is IGR J18135-1751 (aka SWIFT J1813.6-1753; \citealt{Krimm2013, Oh2018}).

\subsection{Spectroscopy} \label{sec:spec}

\begin{figure}
\centering
\includegraphics[width=0.49\textwidth]{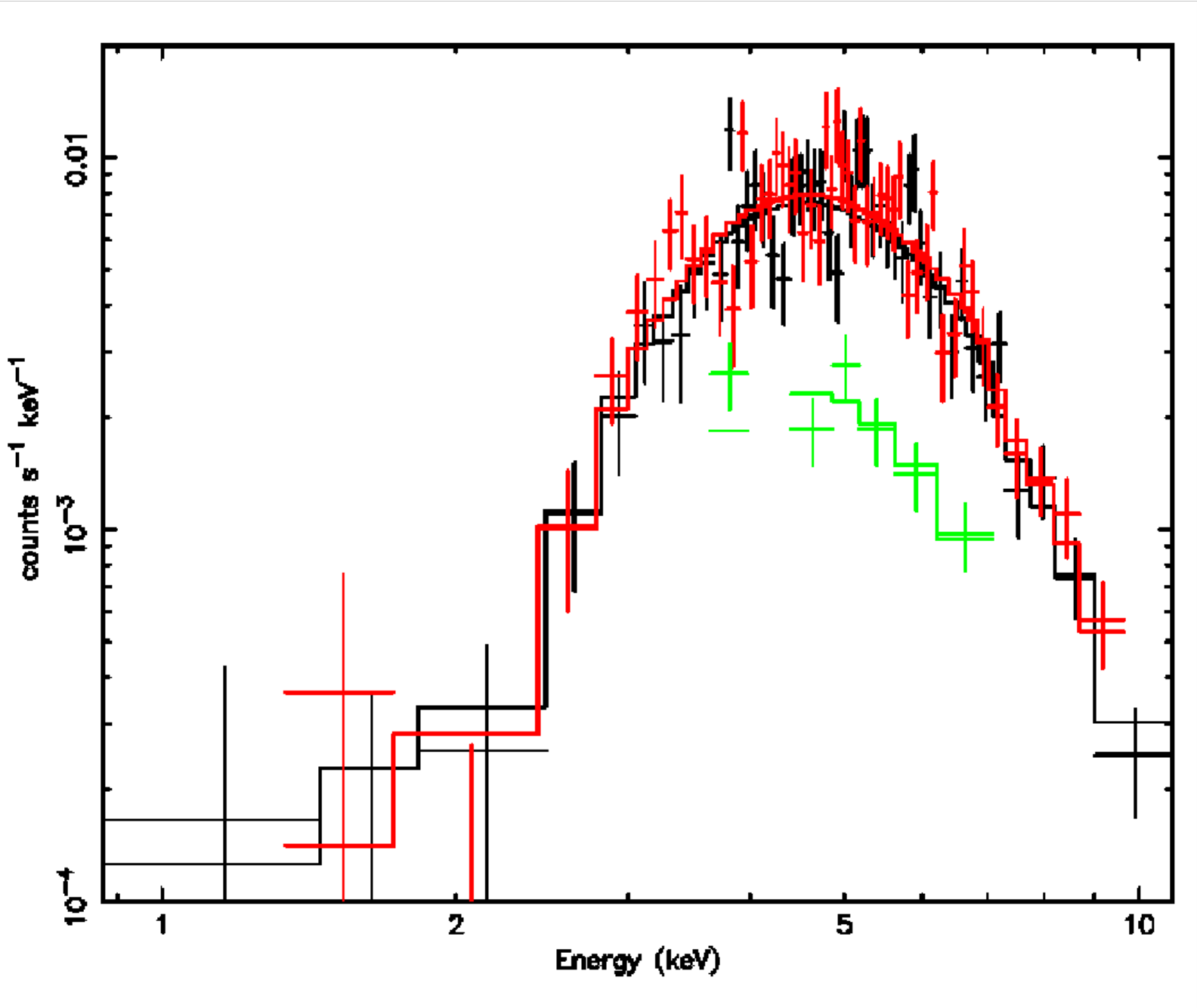}
\caption{Total spectrum of J1813. The black and red points are from MOS1 and MOS2 of {\it XMM-Newton} obs. 0552790101 respectively, while the green ones from the {\it Chandra} obs. 6685. We also report the best-fitted model, as described in Section \ref{sec:spec}.}
\label{fig:spec}
\end{figure}

As described in Section \ref{sec:obs}, we extracted the background-subtracted spectra of J1813 from the {\it XMM-Newton} observation 0552790101 and the {\it Chandra} observation 6685 -- the only datasets that allow for spectral analysis due to sufficient statistics. For the {\it XMM-Newton} observation, we applied a time cut to match the active state of J1813 (T > 354610080), as reported in Section \ref{sec:lc} and shown in Figure \ref{fig:lc_final} (right panel).
For this analysis, we relied on 1004, 1066, and 149 background-subtracted counts from the MOS1 and MOS2 {\it XMM-Newton} cameras and {\it Chandra}, respectively. We allowed the normalization of the different instruments to vary freely to account for cross-calibration differences and, in the case of the {\it Chandra} spectrum, the differing overall flux.\\
We obtained a good fit using the absorbed power-law model {\tt tbabs*powerlaw} \citep[abundances and cross sections respectively from][]{wil00,ver96}, with a $\chi^2$ of 106.74 for 100 degrees of freedom, resulting in a nhp=0.30. The best-fitted values are N$_H$ =  (1.78$\pm$0.15)$\times10^{23}$ cm$^{-2}$ and $\Gamma$ = 1.66$\pm$0.20 (1$\sigma$ errors). The absorbed 0.2-12 keV {\it XMM-Newton} flux is (2.10$_{-0.33}^{+0.02}$)$\times10^{-12}$ erg cm$^{-2}$ s$^{-1}$, that translates into an unabsorbed flux of (7.19$_{-1.13}^{+0.07}$)$\times10^{-12}$ erg cm$^{-2}$ s$^{-1}$. Similarly, the absorbed 0.3-10 {\it Chandra} flux is (2.55$_{-0.64}^{+0.17}$)$\times10^{-13}$ erg cm$^{-2}$ s$^{-1}$, that translates into an unabsorbed flux of (1.38$_{-0.35}^{+0.09}$)$\times10^{-12}$ erg cm$^{-2}$ s$^{-1}$.

To search for possible spectral variations during the active phase over time, we used the {\it XMM-Newton} data to study the variation of the hardness ratio (HR), defined as the counts in a hard energy band over the counts in a soft energy band, as a function of time. As a first attempt, we produced standard background-subtracted light curves in the 2-5 keV (soft band) and 5-12 keV (hard band) energy bands (pattern 0-12) from the MOS1 and MOS2 cameras. We created an HR curve as the sum of the counts from both cameras in the hard band divided by the sum of the counts in the soft band, with appropriate error propagation. We used a bin size of 1 ks to achieve acceptable $\chi^2$ statistics and reasonable error bars. A fit with a constant yielded an acceptable result, with $\chi^2 = 54.3$, 36 degrees of freedom (dof), and nhp = 0.025.\footnote{We note that, by using the definition of HR as $(cts_H-cts_S)/(cts_H+cts_S)$, where $cts$ are the photons in the hard (5-12 keV) and soft band (2-5 keV) for a given time bin, we obtain a consistent result.}
Next, we produced a similar HR curve using the flux curves from EXTraS in the bands 2-4.5 and 4.5-12 keV (EXTraS can only utilize the standard {\it XMM-Newton} energy bands, as defined in the 3XMM catalog). Again, we obtained an acceptable constant fit, with $\chi^2 = 52.6$, 36 dof, and nhp = 0.037. Thus, we do not observe any appreciable variation in J1813’s spectrum over time.\\
We also searched for possible spectral variations during the active phase of the {\it XMM-Newton} observation as a function of flux. For each MOS camera we extracted two spectra based on the cuts described above and an additional temporal cut: the "quiet" spectrum was taken when the EXTraS flux curve described in Section \ref{sec:lc} (Figure \ref{fig:lc_final}) had a value below 4$\times10^{-12}$ erg cm$^{-2}$ s$^{-1}$, while the "flaring" spectrum was taken when it exceeded that threshold. This limit was chosen to approximately equalize the counts for the two states; moreover, the "quiet" spectrum primarily comprises the possible "non-flaring" state discussed earlier (shown in blue in Figure \ref{fig:lc_final}). A simultaneous fit of all four spectra using the model described above, with all parameters except for the normalizations linked, resulted in an acceptable fit with $\chi^2 = 100.8$, 95 dof, and nhp = 0.185.
We note that a similar approach—using spectra for each camera that only included the possible "non-flaring" state (the blue period shown in Figure \ref{fig:lc_final}) and spectra that included the remaining active state—also yielded an acceptable fit, with $\chi^2 = 105.6$, 89 dof, and nhp = 0.111. In this case, the "non-flaring" spectra had a total of less than 200 counts, leading to a poorly constrained fit.
In conclusion, we find no appreciable variation in J1813’s spectrum with flux nor with time.

\section{Optical to mid-infrared data} \label{sec:opt}

\begin{figure}[ht!]
  \centering
 \includegraphics[width=0.49\textwidth]{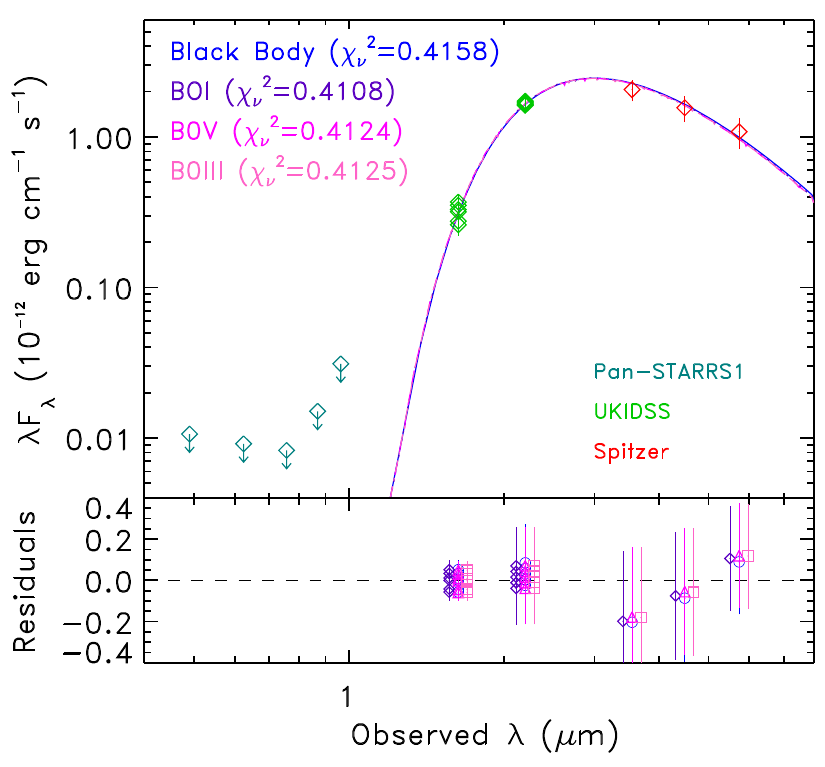}
 \caption{Optical-to-MIR spectral energy distribution of J1813 (open diamonds) from Pan-STARRS1 (teal), UKIDSS (green), and Spitzer (red). The curves represent the best-fit models, an absorbed black-body (blue solid line), and three absorbed stellar models from the \citet{kurucz93} template library, a B0I model (dotted purple line), B0V model (dashed magenta line), and a B0III model (dot-dashed pink line). The residual (observed minus predicted) fluxes are shown with a small horizontal shift for clarity in the bottom panel (black-body: blue circles; B0I: purple diamonds, B0V: magenta triangles, B0III: pink squares). All three models assume an \av=38\,mag. }
\label{fig:sed}
\end{figure}

To identify J1813 it is crucial to determine the nature of the companion star. 
To this end, we searched for the optical and infrared (IR) counterpart of J1813 in public databases \citep[i.e., in the VizieR catalog library;][]{vizier}. The source was observed, although not detected, in the optical by the Panoramic Survey Telescope \& Rapid Response System \citep[Pan-STARRS or PS1;][]{chambers16,flewelling20} telescope. Based on the PS1 survey depth we assign upper limits to the optical emission of J1813 (i.e., the AB magnitude 5$\sigma$ limits are 23.3, 23.2, 23.1, 22.3, 21.4 in the $g$, $r$, $i$, $z$, and $y$ bands, respectively). Multiple detections in H and K bands are available from the UKIRT Infrared Deep Sky Survey \citep[UKIDSS;][]{lawrence07} which reveal some level of variability in the H band. Mid-IR data are available from the Galactic Legacy Infrared Midplane Survey Extraordinaire \citep[GLIMPSE;][]{glimpse} Legacy Program carried out with the {\it Spitzer Space Telescope}. All the available data are listed in Table~\ref{tab:data} and shown in Fig.~\ref{fig:sed}.

\begin{table} 
\caption{\label{tab:data} Near and mid-IR measurements of J1813.} 
\centering 
\begin{tabular}{c c c c}
\hline\hline
 Band        &   $\alpha$      & $\delta$         &   $S_{\nu}$    \\
             &   h:min:s       & \deg:\arcmin:\arcsec\ &  mJy      \\
\hline
\multicolumn{4}{c}{UKIDSS}\\
  H           &  18:13:30.1774  & $-$17:51:10.357  &  0.150$\pm$0.007 \\
  H           &  18:13:30.1774  & $-$17:51:10.357  &  0.201$\pm$0.005 \\
  H           &  18:13:30.1774  & $-$17:51:10.357  &  0.179$\pm$0.005 \\
  H           &  18:13:30.1759  & $-$17:51:10.328  &  0.142$\pm$0.008 \\
  H           &  18:13:30.1759  & $-$17:51:10.328  &  0.191$\pm$0.005 \\
  H           &  18:13:30.1759  & $-$17:51:10.328  &  0.173$\pm$0.006 \\
  K           &  18:13:30.1774  & $-$17:51:10.357  &  1.21$\pm$0.01 \\
  K           &  18:13:30.1774  & $-$17:51:10.357  &  1.27$\pm$0.01 \\
  K           &  18:13:30.1774  & $-$17:51:10.357  &  1.25$\pm$0.01 \\
  K           &  18:13:30.1759  & $-$17:51:10.328  &  1.19$\pm$0.01 \\
  K           &  18:13:30.1759  & $-$17:51:10.328  &  1.21$\pm$0.01 \\
  K           &  18:13:30.1759  & $-$17:51:10.328  &  1.23$\pm$0.01 \\
 \multicolumn{4}{c}{GLIMPSE}\\
 3.6$\mu$m    &  18:13:30.1829  & $-$17:51:10.494  &  2.44$\pm$0.16 \\
 4.5$\mu$m    &  18:13:30.1829  & $-$17:51:10.494  &  2.34$\pm$0.23 \\
 5.8$\mu$m    &  18:13:30.1829  & $-$17:51:10.494  &  2.07$\pm$0.28 \\
\hline
\end{tabular}\\
\end{table}

The optical-to-MIR spectral energy distribution (SED) of J1813 is extremely red, implying significant obscuration. We fitted the SED with an absorbed black body following \citet{rahoui08}. The black body model can be parameterized as:
\begin{equation}
\lambda F_{\lambda} = \frac{2\pi h c^2}{D_{\ast}\lambda^4}10^{-0.4A_{\lambda}} \frac{R_{\ast}^2}{e^{\frac{h c}{\lambda K T_{\ast}}}-1}
\end{equation}
where $T_{\ast}$, $R_{\ast}$ , and $D_{\ast}$ are, respectively, the companion star black body temperature, radius, and distance. The free parameters are $T_{\ast}$, $R_{\ast}/D_{\ast}$, and the extinction \av. We assume as extinction curve a combination of laws derived for the Galactic plane in different wavelength ranges \citep{cardelli89,indebetouw05,chiar06}. The best fit parameters are determined through $\chi^2$ minimization after adding a systematic 10\% uncertainty to the flux measurements in quadrature. The best fit black body model and the derived reduced-$\chi^2$ are displayed in Fig.~\ref{fig:sed}. A single black body yields a good fit with $T_{\ast}=32000\pm4000$\,K, \av=38$\pm$1, and $R_{\ast}/D_{\ast}=(3.5\pm0.5)\times 10^{-11}$. Based on the estimated stellar temperature, the stellar class corresponds to a B star \citep{payne25}. The stellar class interpretation is confirmed by the best-fit stellar model obtained after fitting the SED with the \citet{kurucz93} stellar library. The best fit stellar models, also shown in Fig.~\ref{fig:sed}, are given by the class B0I, B0V and B0III stellar models absorbed by an extinction consistent with that derived from the black body model.
Assuming a typical radius of a B0 star ($R_{\ast}\simeq 10-20\,R_{\odot}$; e.g. \citealt{coleiro13a}), with the lower radius more in line with a main sequence star and the larger radius with a giant, the estimated distance of J1813 would be $D_{\ast}\simeq 6.5-13.1$\,kpc.
The derived visual absolute magnitude of J1813 from the black body model ranges from  $-$6.8 to $-$8.4, depending on the assumed distance $D_{\ast}$. 
Based on the derived extinction, temperature and luminosity, we conclude that the companion star of J1813 must be an extremely absorbed B star at a distance of 7--13\,kpc.  
A HMXB showing 
a similar high extinction value is IGR\,J16320$-$4751, an O/B supergiant whose compact object is a NS \citep{lutovinov05} and with an estimated extinction is \av=35.4\,mag \citep{rahoui08}. This is quite large compared to the
extinction values \av$\lesssim$20 usually shown by these sources \citep{pellizza06,rahoui08,torrejon10,coleiro13a,coleiro13b,hare19,fortin20,sidoli22}. 

Assuming the empirical relation between the hydrogen column density and the visual extinction as derived by ~\citet{foight16} (i.e., \nh (cm$^{-2}) = (2.87 \pm 0.12) \times 10^{21}$ \av), the extinction derived for J1813 corresponds to \nh$\simeq (1.1 \pm 0.1) \times 10^{23}$\,cm$^{-2}$. This column density is slightly lower than the value derived from the X-ray spectrum, which is typical pf highly absorbed HMXB systems \citep[e.g.,][]{rahoui08}. A similar difference is observed in IGR\,J16320$-$4751, for which the estimated extinction corresponds to a hydrogen column density \nh$=1.0 \times 10^{23}$\,cm$^{-2}$, lower than  that derived from the X-ray spectrum is \nh$\sim 2.1\times 10^{23}$\,cm$^{-2}$.

\section{Discussion} \label{sec:disc}

We report here on an X-ray transient source showing flaring behavior we have identified within EXTraS, analysing {\it XMM-Newton} archival data.
During a 100 ks-long observation, the source remained undetected during the initial 60 ks followed by an active phase with several X-ray flares (lasting a few ks each).

A candidate optical counterpart was found within the X-ray error region. Its SED resutls to be consistent with that of a B0 star with a high extinction, implying that J1813 is a new HMXB. From the fit results we have found that any stellar luminosity class is in agreement with the observed SED, although a slightly better fit is obtained with a B0I star.
Therefore the optical counterpart alone (which can be a main sequence, a giant or a supergiant B star) indicates only a membership within the broad family of HMXBs, but not a secure sub-type nature. The optical identification indicates a source located at a distance of $\sim$10 kpc (in-between 7 and 13 kpc), allowing us to calculate an average X-ray luminosity of the active phase of about 10$^{35}$~erg~s$^{-1}$.

The transient nature of the X-ray emission suggests either a BeXRB or a SFXT. The shape of its X-ray light curve, the presence of many short flares, their temporal and spectral properties strongly favor a SFXT. 
In particular, the light curve of J1813 is similar to that observed in the archetypal SFXT IGR\,J17544-2619 with {\it XMM-Newton} \citep{Bozzo2016} and {\it Chandra} \citep{Zand2005}: a 150 ks long  {\it  XMM-Newton} observation caught IGR\,J17544$-$2619 in an initial dim X-ray state, followed by a bright X-ray flaring activity lasting 7~ks \citep{Bozzo2016}, very similar to the light curve we have discovered from J1813. A FRED-like profile was observed in the same SFXT  IGR\,J17544$-$2619 with {\it Chandra} \citep{Zand2005}. 

The X-ray luminosity of the active phase in J1813 indicates flaring emission in an intermediate X-ray luminosity, frequently observed in SFXTs \citep{Sidoli2008, Romano2015, Sidoli2019}.
The amplitude of the observed long-term X-ray flux variability is $>$500, measured from the peak of the brightest flare observed by {\it XMM-Newton} (n.~8 in Table~\ref{tab:lcpar}) to the most stringent upper limit reported in Section \ref{sec:lc}.
The implied X-ray luminosities are in the range from L$_{X}$$<$8$\times$10$^{32}$~erg~s$^{-1}$ to 4$\times$10$^{35}$~erg~s$^{-1}$ 
(2-12 keV, assuming a distance of 10 kpc), well within the typical SFXT flares' properties \citep{Sidoli2019}.
The time averaged spectrum during the X-ray flaring activity is highly absorbed (N$_H$ in excess of 10$^{23}$ cm$^{-2}$), well fitted by a power-law model ($\Gamma$$\sim$1.7), emitting an average luminosity of
7.1$\times$10$^{34}$~erg~s$^{-1}$ (0.2-12 keV). 
Power-law-like spectra with similar slopes are typical of SFXTs at this level of X-ray emission \citep{Sidoli2008}, while much flatter spectra ($\Gamma\sim0.0-1.0$) are observed only during SFXT outbursts.

The temporal parameters of J1813's flares (Table~\ref{tab:lcpar}) match those reported in the investigation of SFXTs observed by {\it XMM-Newton} and analysed within EXTraS \citep{Sidoli2019}. In particular, the waiting time (i.e. the interval of time between two consecutive flare peaks) is in the range from 0.69 to 12.6 ks, in agreement with the waiting time measured in SFXTs (0.15--12.1 ks, with a distribution peaking around 0.4--5 ks). The rise and decay times overlap with the values measured in SFXTs as well. 
The lack of any known X-ray outbursts (above 10$^{36}$~erg~s$^{-1}$) in archival X-ray observation of J1813 is also in line with the typical very low duty cycle of SFXTs (less than 5\%, but it can be much lower in some members of the class; \citealt{Sidoli2018}).

For what concerns the source duty cycle, we have estimated that J1813 spends approximately 40\% of the time above an X-ray flux of a few 10$^{-13}$~erg~cm$^{-2}$~s$^{-1}$ (Sect.~\ref{sec:lc}), i.e. above a luminosity of a few 10$^{33}$~erg~s$^{-1}$. This is in agreement with what measured by {\it Swift}/XRT monitoring of a sample of SFXTs \citep{Romano2015}, in particular the sources IGR~J08408–4503, XTE~J1739–302 and IGR~J17544–2619. 

On the other hand, if we compare J1813 properties with BeXRBs, short X-ray flares are very rarely observed in this kind of HMXBs. And when present, they happen just before or after the peak of an outburst, with flare peaks much more luminous than those observed in J1813 \citep{Reig2008}. 
Two anomalous, isolated flares were caught recently by eRosita (0.2-10 keV) from the BeXRB  A0538-66 \citep{Ducci2022}, at orbital phases far from the periastron passage and without concomitant outbursts. However, these flares showed a larger X-ray luminosity (a few 10$^{36}$~erg~s$^{-1}$) than in J1813 and an unconstrained duration (in-between 42 s and 29 ks the first flare, and in-between 29 ks and 57 ks the second one).

In conclusion, the observed properties of this new HMXB strongly support an identification with an SFXT. However, optical and near infrared spectroscopy is needed to confirm the luminosity class of the donor star and reduce the uncertainty in the source distance.

\begin{acknowledgements}
This research has made use of the VizieR catalogue access tool, CDS, Strasbourg Astronomical Observatory, France (DOI : 10.26093\/cds\/vizier).
The Pan-STARRS1 Surveys (PS1) and the PS1 public science archive have been made possible through contributions by the Institute for Astronomy, the University of Hawaii, the Pan-STARRS Project Office, the Max-Planck Society and its participating institutes, the Max Planck Institute for Astronomy, Heidelberg and the Max Planck Institute for Extraterrestrial Physics, Garching, The Johns Hopkins University, Durham University, the University of Edinburgh, the Queen's University Belfast, the Harvard-Smithsonian Center for Astrophysics, the Las Cumbres Observatory Global Telescope Network Incorporated, the National Central University of Taiwan, the Space Telescope Science Institute, the National Aeronautics and Space Administration under Grant No. NNX08AR22G issued through the Planetary Science Division of the NASA Science Mission Directorate, the National Science Foundation Grant No. AST–1238877, the University of Maryland, Eotvos Lorand University (ELTE), the Los Alamos National Laboratory, and the Gordon and Betty Moore Foundation.
This work is based in part on data obtained as part of the UKIRT Infrared Deep Sky Survey.
This work is based in part on observations made with the Spitzer Space Telescope, which was operated by the Jet Propulsion Laboratory, California Institute of Technology, under a contract with NASA.
This research has made use of data obtained from the 4XMM XMM-Newton serendipitous source catalogue compiled by the XMM-Newton Survey Science Centre consortium.
This research has made use of data obtained from the Chandra Data Archive and the Chandra Source Catalog, both provided by the Chandra X-ray Center (CXC).
\end{acknowledgements}

\bibliographystyle{aa}
\bibliography{biblio.bib}

\end{document}